\documentclass[twocolumn,noshowpacs,nofootinbib,10pt]{revtex4}

\usepackage{float,xcolor,upgreek,yfonts,tikz}
\usepackage{color} 
\usepackage{graphicx}
\usepackage{graphicx,float}\usepackage[all]{xy}
\usepackage{amsmath,upgreek}
 \newcommand{\blt}{\textcolor{black}}
\usepackage{amssymb}

\usepackage{alphabeta}
\usepackage{textgreek}
\newcommand{\beq}{\begin{eqnarray}}
\newcommand{\eeq}{\end{eqnarray}}
\newcommand{\be}{\begin{equation}}
\newcommand{\ee}{\end{equation}}

\newcommand{\bea}{\begin{eqnarray}}
\newcommand{\eea}{\end{eqnarray}}

\usepackage{mathrsfs}
\newcommand{\ba}{\begin{eqnarray}}
\newcommand{\ea}{\end{eqnarray}}
\newcommand{\clt}{\textcolor{black}}

\def\NP{c_{\scriptscriptstyle\text{NP}}}

\DeclareFontFamily{OT1}{rsfs}{}
\DeclareFontShape{OT1}{rsfs}{m}{n}{ <-7> rsfs5 <7-10> rsfs7 <10->
rsfs10}{} \DeclareMathAlphabet{\mycal}{OT1}{rsfs}{m}{n}

\usepackage[T3,T1]{fontenc}
\DeclareSymbolFont{tipa}{T3}{cmr}{m}{n}
\DeclareMathAccent{\invbreve}{\mathalpha}{tipa}{16}

\usepackage[colorlinks,hyperindex,unicode]{hyperref}
\definecolor{green1}{RGB}{0,128,0} 
\hypersetup{hidelinks,backref=true,pagebackref=true,hyperindex=true,colorlinks=true,breaklinks=true,urlcolor= blue}
\hypersetup{%
  colorlinks = true,
  linkcolor  = blue,
  citecolor = cyan,
}
\usepackage{bookmark,textgreek}
\usepackage{hyperref,color,xcolor}
\hypersetup{hidelinks,hyperindex=true,colorlinks=true,breaklinks=true,urlcolor= blue}
\hypersetup{%
  colorlinks = true,
  linkcolor  = blue
}
\newcommand\orcidroldao{{\href{https://orcid.org/0000-0003-3978-532X}{\orcidicon}}}
\newcommand\orcidwillians{{\href{https://orcid.org/0000-0001-9750-2637}{\orcidicon}}}
\newcommand{\orcidicon}{%
	\begin{tikzpicture}
	\draw[lime, fill=lime] (0,0)
		circle [radius=0.16]
		node[white] {{\fontfamily{qag}\selectfont \tiny ID}};
	\draw[white, fill=white] (-0.0625,0.095)
		circle [radius=0.007];
	\end{tikzpicture}	\hspace{-2mm}
}

\begin{document}
\title{Differential configurational entropy and the gravitational collapse of a kink}

\author{W. Barreto\orcidwillians\!\!}
\affiliation{Federal University of ABC, Center of Natural Sciences, Santo Andr\'e, 09580-210, Brazil}\email{willians.barreto@ufabc.edu.br}
\affiliation{Centro de F\'{i}sica Fundamental, Universidad de Los Andes, M\'{e}rida 5101, Venezuela}
\author{R. da Rocha\orcidroldao\!\!}
\affiliation{Federal University of ABC, Center of Mathematics,  Santo Andr\'e, 09210-580, Brazil.}
\email{roldao.rocha@ufabc.edu.br}

\begin{abstract}
The gravitational instability of kinks is scrutinized in the context of information entropy, including the black hole formation near the scalar field critical collapse. \clt{The differential configurational entropy (DCE) is computed and examined for globally perturbed static kinks and the far-from-equilibrium kink solutions as well. In the case of the first kink, the DCE determines a critical value of the perturbation parameter regulating points of bifurcation that characterize a phase transition, whose supercritical range leads to collapse and subsequent black hole formation. For the far-from-equilibrium kink solutions, the DCE supports black hole formation.}
\end{abstract}

\maketitle

\section{Introduction}
The configurational entropy (CE) quantifies information and its transmission rate across channels of communication. 
The CE computes the amount of information that is concealed in stochastic variables and has been successfully implemented to survey a plethora of random physical systems and to scrutinize and quantify their different and relevant properties. In this way entropy in statistical mechanics and information entropy are interrelated concepts \cite{Shannon:1948zz}. Similar to the Gibbs entropy, which generalizes the Boltzmann entropy for microstates that are not equiprobable, information entropy expresses the alterations of arrangement and disorder immanent to a given physical system.  Shannon's information entropy quantifies how much information is necessary to describe the specific detailed microscopic configuration of a system, once the associated macrostate is considered. Also, the rate of compression of the information is addressed by Shannon's formulation.  
\blt{The CE was at first presented by Gleiser and Stamatopolous as a quantity that compares exact soliton-type solutions to approximations of it,  eliminating the degeneracy of equal-energy ans\"atze regarding models containing a 
scalar field \cite{gs12a}. Besides, other configurational information measures  were proposed to evaluate information and shape complexity of physical systems appearing in field theories 
\cite{Gleiser:2018kbq,Gleiser:2018jpd}. Several 
non-linear scalar field models that include kinks, solitons, bounces, and bubbles were scrutinized from the point of view of the CE, characterizing first-order phase transitions. Ref. \cite{gs12a} was the first seminal result in the literature to demonstrate that the lower the energy of a trial scalar field estimating exact solutions in field theory, the lower its associated CE is. In this setup, the CE was shown to pick  the best fit for scalar field solutions. Matter configurations, encompassing kinks, solitons, oscillons, and stellar distributions, have been studied from the point of view of the DCE, with important information about  stability   \cite{Gleiser:2018jpd,gs12b}.  Ref. \cite{Sowinski:2015cfa} analyzed the CE profile throughout a second-order phase transition, with an approach to criticality resembling the scaling behavior underlying the Kolmogorov-type turbulence in fluid flows, emulating the important concept of informational turbulence \cite{Sowinski:2016vxz}}. When continuous physical systems are taken into account, for low limits of almost lossless data compression, the differential configurational entropy (DCE) is the most suitable measure of information entropy \cite{Gleiser:2018kbq,Gleiser:2018jpd}. It can be evaluated by taking any spatially-localized, $\ell^2$-norm integrable scalar field \cite{gs12a,gs12b}. 
Global critical points of the DCE comply with microstates that have maximal configurational stability, being thus the preferred state of occupation of the system under scrutiny. 
The DCE has been applied to the solution of a variety of relevant physical systems. 
Gauge/gravity dualities and phase transitions were successfully investigated with the tools of the DCE \cite{Bernardini:2016hvx,Braga:2017fsb,Braga:2020hhs,Braga:2020myi,Braga:2020opg}. 
Also, the DCE provided new features of black holes in anti-de Sitter (AdS) space \cite{Braga:2019jqg,Lee:2017ero}. The Hawking--Page first-order phase transition was shown to correspond to the DCE global critical point \cite{Braga:2016wzx}. In this respect, although AdS black holes can share stable thermal equilibrium with radiation, they are not the preferred state below a certain critical temperature, when thermal AdS emerges as the prevalent  contribution to the partition function in AdS/CFT \cite{Lee:2021rag}. The stability 
of compact stellar distributions was also addressed in Ref. \cite{Fernandes-Silva:2019fez}, corresponding to global minima of the DCE computed from the star energy density.
The DCE has been auspiciously
employed to address the stability of spatially-localized configurations of scalar fields in several different physical contexts. The DCE was used to derive the Chandrasekhar critical density of stellar core remnants consisting of electron-degenerate matter in Ref. \cite{Gleiser:2013mga}, of collapsed neutron cores of massive supergiant stars in Ref. \cite{Gleiser:2015rwa}, and of Bose--Einstein condensates 
of gravitons, in Ref. \cite{Casadio:2016aum}. 
Also, stellar configurations composed by gravitons in AdS were perused from the point of view of the DCE, derived as solutions of Einstein--Hilbert gravity with a gauge field, backreacting to the spontaneous compactification \cite{dr21}. The DCE was also investigated in other different phenomenological contexts, with relevant results 
\cite{Karapetyan:2018yhm,Karapetyan:2018oye,Bazeia:2018uyg,Correa:2015lla,Bazeia:2021stz,Correa:2016pgr,Cruz:2019kwh,Lima:2021noh}.

The unequivocal experimental detection of gravitational waves, which are emitted from coalescing binary systems of stars and black holes as well, has currently consisted of prominent directions in physics  \cite{LIGOScientific:2017vwq,Xue:2019nlf}. 
In the strong gravitational field regime, gravity can be successfully described by the theory of general relativity and some relevant generalizations, which can be probed by current observations, mostly at eLISA and LIGO.  
The gravitational critical behavior has been thoroughly studied in several contexts, employing a variety of theoretical and computational methods \cite{c93,Barreto:2017kta}. This universal phenomenon appears in a myriad of circumstances, in particular including the gravitational collapse of scalar fields in several types of spacetimes 
\cite{Santos-Olivan:2015yok}. A case of relevance involves a spherically symmetric scalar field, in the strong gravitational field limit approaching collapse and subsequent  black
hole formation. 
The gravitational critical behavior becomes even more fascinating and relevant when a mass gap is present in the  evolution of a scalar field kink
\cite{bglw96}, therefore revealing important features of the
scalar field gravitational collapse. Kinks have been explored in various contexts, including new gravitational sectors \cite{Bazeia:2012rc,Bazeia:2014hja}.
Ref. \cite{Santos-Olivan:2015yok} addressed critical points of bifurcation, determining the ramification into either a mass gap or no mass gap. An efficient procedure to approach this theme is to assume the Einstein--Klein--Gordon coupled system of effective field equations, to look over the gravitational collapse and subsequent black hole formation and its dynamics.
The Einstein--Klein--Gordon system, coupling gravity to a scalar
field, implements a generalization of general relativity and can solve observed deviations. Some conditions may also circumvent the no-hair conjecture \cite{Ovalle:2020kpd,Ovalle:2021jzf}. 
With it in hand, the critical behavior of gravitational
collapse can be better probed with data regarding coalescing  binary stars and black hole  binary
systems \cite{Pretorius:2005gq}, with their particular experimental signatures detected from the gravitational collapse into gravitational wave radiation. Several open problems,  regarding the stability of the gravitational collapse of Einstein--Klein--Gordon solutions, can be addressed using the DCE apparatus. 
\clt{Here we consider the gravitational collapse of a kink as studied in Refs. \cite{bglw96,bcdrr16}, to explore the DCE under full numerical spacetimes with nontrivial features, such as 
the critical behavior discovered by Choptuik \cite{c93}. We set initially two kinds of kinks; one close to the well-known Janis--Newman--Winicour static solution \cite{jnw68} with a cutoff of the singular region and another very far--from--equilibrium. The first exhibits the critical behavior with a mass gap \cite{bcdrr16},
and the second always forms a black hole \cite{bglw96}. The idea is to use these two extreme situations to explore how the DCE catches up with these highly nonlinear dynamical processes for the gravitational collapse and try to establish once more its physical robustness}, \clt{also corroborating to well-established results in the literature as well as prospecting new relevant features regarding phase transitions and the gravitational collapse.}

\clt{This paper is organized as follows: Section \ref{17} is devoted to studying the Einstein--Klein--Gordon field equations, stressing the (matter) fluid correspondence with the massless scalar field. For it, the Bondi--Sachs metric plays a prominent role in defining a Bondi frame, and the Bondi mass is addressed. In Section \ref{2s}, the energy density associated with the scalar field is  constructed upon the spatially-localized scalar field, its Fourier transform, the modal fraction, and the DCE, are discussed. Some previous results for the critical behavior are summarized and we give some information about the used numerical solver and scripting. In Section \ref{3s}two initial conditions for the considered evolutions are set, including  globally perturbed static kinks and the far-from-equilibrium kink solutions.  In Section \ref{4s} the results are analyzed and discussed, investigating the role of the DCE in determining a phase transition. It  distinguishes two main types of  evolution profiles into final equilibria, that can either lead to a final black hole state or a static solution. Ulterior features and analysis are reported, also in Section \ref{conclu}.}

\section{Field equations}
\label{17}
We consider the Einstein-Klein-Gordon equations of motion
\begin{subequations}
\beq
G_{\mu\nu}&=&-8\pi T_{\mu\nu},\label{fe1}\\
\square \Upphi &=& 0,\label{fe2}
\eeq
\end{subequations}
where $G_{\mu\nu}$ denotes the Einstein's tensor
\begin{equation}
T^\Upphi_{\mu\nu}=\nabla_\mu\Upphi\nabla_\nu\Upphi -\frac{1}{2}
g_{\mu\nu}\nabla^\alpha\Upphi\nabla_\alpha\Upphi.
\end{equation}
Spherically symmetric solutions have an initial value setup that comprises outgoing null cones emerging out a central geodesic. For it, one denotes by $u$ the (proper) time coordinate along the geodesic, assuming constant values on the outgoing null cones. 
Therefore the Bondi--Sachs coordinates, $x^\mu=(u, r, \theta,\phi)$ can be employed to write the Bondi--Sachs metric under spherical symmetry \cite{b64}
\begin{equation}
ds^2=e^{2\beta}\left(\frac{V}{r}du^2+2dudr\right)-r^2(d\theta^2+\sin^2\theta d\phi^2), \label{bm}
\end{equation}
It is worth mentioning that the normal covector field $k_\mu = \partial_\mu u$ is null, also implying that the corresponding future-pointing vector field $k^\mu = -g^{\mu\nu}\partial_\nu u$ is a tangent vector field along the null-ray manifold. 
When $r\to0$, the conditions 
\begin{subequations}
\beq
V(u,r)&=&r+\mathcal{O}(r^3),\\
\beta(u,r)&=&\mathcal{O}(r^2),
\eeq
\end{subequations}
must be imposed for the Minkowski spacetime be recovered.

The Misner--Sharp mass function,
$\tilde m(u,r)$ \cite{ms64} introduced by means of
\begin{equation} 
\tilde m=\frac{1}{2}(r-Ve^{-2\beta}),\label{5} 
\end{equation}
measures the energy content in the sphere of radius $r$ and it reduces to the
Arnowitt--Deser--Misner and Bondi masses in the appropriate limits.

The energy-momentum tensor for the massless scalar field minimally coupled to gravity can be characterized as a radiating and anisotropic fluid \cite{hs97,bbc10}
\beq
T^\Upphi_{\mu \nu}&=&(\rho^\Upphi+p_t^\Upphi)u_{\mu}u_{\nu}+\varepsilon^\Upphi
 l_{\mu}l_{\nu}-p_t^\Upphi g_{\mu \nu}\nonumber\\&&\qquad+(p^\Upphi_r-p_t^\Upphi)
\chi_{\mu}\chi_{\nu},
\eeq
with $u^\mu u_\mu=1,$
$l^{\mu}l_{\mu}=0,$
$\chi^\mu \chi_\mu=-1,$
when one identifies the 4-velocity for an observer at rest,  with respect to the metric (\ref{bm}), by 
\begin{equation}
u^\mu=\left(1-\frac{2\tilde{m}}{r}\right)^{-1/2}e^{-2\beta}\delta^\alpha_0, \label{4v}
\end{equation}
and the null and the space-like vectors respectively expressed as
\begin{subequations}
\beq l_\mu&=&\left(1-\frac{2\tilde{m}}{r}\right)^{1/2}e^{2\beta}\delta^0_\mu,\\
\chi_\mu&=&\left(1-\frac{2\tilde{m}}{r}\right)^{-1/2}\delta^1_\mu.\eeq
\end{subequations}
The scalar energy flux, the scalar energy density, the scalar radial pressure, and
the scalar tangential pressure,  are respectively given by 
\beq
\!\!\!\!\epsilon^\Upphi&=&e^{-2\beta}\left[e^{-2\beta}\left(1-\frac{2\tilde{m}}{r}\right)^{-1}(\Upphi_{,u})^2 -\Upphi_{,u}\Upphi_{,r}\right],
\label{rad_phi}\\
\!\!\!\!\rho^\Upphi &=& p^\Upphi_r = \frac 1 2 \left(1-\frac{2\tilde{m}}{r}\right)(\Upphi_{,r})^2,\label{rofi}\\
\!\!\!\!p_t^\Upphi&=&\Upphi_{,u}\Upphi_{,r}e^{-2\beta}-p_r^\Upphi,\label{tan_phi}
\eeq
where the comma denotes partial differentiation with respect to the indicated coordinate.
Note that Bondian observers can be purely Lagrangian when 
we deal only with radiation. The observer with four-velocity (\ref{4v}) is resting
at infinity \cite{hbcd07}.

The field equations (\ref{fe1}, \ref{fe2}) reduce to the well known hypersurface equations  
for a scalar field \cite{gw92,bglw96,bcdrr16}
\begin{subequations}
\beq
\beta_{,r}=2\pi r (\Upphi_{,r})^2,\label{221}\\
V_{,r}=e^{2\beta},\label{222}
\eeq
\end{subequations}
whereas the Klein--Gordon equation can be expressed as
and
\begin{equation}
2(r\Upphi)_{,ur}=r^{-1}[r e^{2\beta}(r-2\tilde m)\Upphi_{,r}]_{,r}, \label{we}
\end{equation}
in Bondi coordinates. 
The initial null data that is necessary for analyzing the evolution of the system is given by $\Upphi(0,r)$, in the region exterior to $R$. Also, at this boundary the condition  $\Upphi(u,R)=A=\;$constant is adopted, together with the gauge condition limit $\lim_{r\to\infty}\Upphi(u,r)=0$ and 
$\beta(u,R)=0$. Since the metric must match a
flat inner region, $r<R$, then  $V(u,R)=R$. 

These constraints on the scalar field and the metric as well have therefore a unique evolution across the future. The outcome metric is not asymptotically Minkowski at the future null infinity, being specified by the function $H(u)=\lim_{r\to\infty}\beta(u,r)$, whose additional role is to correlate the Bondi time, $u_{\scriptscriptstyle{B}}$, at the  future null infinity to the proper time by the following expression, 
\be
\frac{du_{\scriptscriptstyle{B}}}{du}=e^{2H(u)}.\label{hu}
\ee
The 4-tuple of coordinates $(u_{\scriptscriptstyle{B}}, r, \theta, \phi$) comprise a  Bondi frame with metric given by Eq. (\ref{bm}), as long as 
the concomitant mappings $V \mapsto V_{\scriptscriptstyle{B}} = e^{-2H}V$ and $\beta\mapsto\beta_{\scriptscriptstyle{B}} =
\beta - H$ are implemented. Using the Bondi time is suitable to  probe asymptotic quantities, whereas the proper time is appropriate to study event horizons \cite{bcdrr16}.
The Bondi mass of the Einstein--Klein--Gordon system reads \cite{gw92}
\beq
M(u)&=&2\pi\int_R^\infty e^{2H(u)-\beta(u,r)}r^2(\Upphi_{,r})^2\,dr, \label{bm1}
\eeq
or, equivalently, 
\beq
M&=&2\pi\int_R^\infty rVe^{-2\beta}(\Upphi_{,r})^2\,dr. \label{bm2}
\eeq
\blt{To implement observational aspects of the results here to be presented,  one can make an expansion of the  scalar field $\Upphi$, near the future null infinity, as  \cite{purrer}
\begin{equation}\label{phi_asm}
\phi(u,r) = \frac{c(u)}{r} + \frac{\NP}{r^2} + O(r^{-3}), \end{equation}
where the coefficient $\NP$ of the $1/r^2$-term 
in the expansion is the Newman--Penrose constant of the scalar field.
Replacing Eq. \eqref{phi_asm} into the hypersurface equations
 yields
\beq
\!\!\!\!\!\!\!\!\!\!\!\beta(u,r) \!&\!=\!&\! H(u) - \frac{\pi c^2(u)}{r^2} + O(r^{-3}),\\
\!\!\!\!\!\!\!\!\!\!\!V(u,r) \!&\!=\!&\! e^{2H(u)} \left(r \!-\! 2M(u) \!+\! \frac{\pi c^2(u)}{r}\right) \!+\! O(r^{-2}),
\eeq
where  $H(u)$ is given by Eq. (\ref{hu}).
The Bondi mass-loss equation asserts that 
\begin{equation}\label{eq:mass-loss}
\frac{dM}{du_{\scriptscriptstyle{B}}} = -4\pi \left(\frac{dc(u_{\scriptscriptstyle{B}})}{du_{\scriptscriptstyle{B}}}\right)^2.
\end{equation}
One can therefore analyze radiation emission for near-critical evolution profiles. The monopole moment of the scalar field $c(u_{\scriptscriptstyle{B}})$ in \eqref{phi_asm} has usually  a (damped) oscillation with a frequency that exponentially decreases. Also,  half-periods in $c(u_{\scriptscriptstyle{B}})$ comply with the ones emerging from the slightest quasinormal mode of Schwarzschild-type solutions, with effective background mass $M_1(u_{\scriptscriptstyle{B}})$, obtained when one evaluates the Bondi mass at its inflection points.
Perturbation fields outside black holes evanesce according to an 
inverse power-law tail, as a byproduct of late-time decay of radiative fields, and reflects the backscattering phenomenon of waves modes off the effective curvature potential.  
For late times, perturbation theory yields the field to fall off as $\Upphi\propto u_{\scriptscriptstyle{B}}^{-2}$, near the future null infinity, and $\Upphi\propto u_{\scriptscriptstyle{B}}^{-3}$, near the timelike infinity.
One can consider an observer at constant $r$ and Bondi time $\Delta u_{\scriptscriptstyle{B}}$, elapsed when the emitted radiation pulse  has been detected by the observer \cite{purrer}. The null infinity can be estimated  by the range $\Delta u_{\scriptscriptstyle{B}}\ll r$, defining the so-called astrophysical zone.
Also, the convergence of the perturbation expansion 
requires $\Delta u_{\scriptscriptstyle{B}} \gg M_1$, whereas the mass associated with the effective potential has an upper bound ruled by the mass producing the initial ingoing pulse, $M_{\scalebox{.66}{\textsc{init}}}$, where $\Delta u_{\scriptscriptstyle{B}} \gg M_{\scalebox{.66}{\textsc{init}}}$.
One may observe tails for $\Delta u_{\scriptscriptstyle{B}} \gtrsim 10^3 M_{\scalebox{.66}{\textsc{init}}}$.} 


\section{DCE and critical gravitational collapse of a kink}
 \label{2s}
Several relevant measures of information entropy have been implemented, mainly in the last decade, and utilized to study gravitational systems and quantum field theories. Among them, the configurational entropy (CE) and its differential version (DCE) play a prominent role in investigating several physical systems and determining their preferred occupied state, always corresponding to critical points of the CE and the DCE. 
The CE can be established, once one regards the original formulation of information entropy by Shannon \cite{Shannon:1948zz}, which defined information by the expression
$
S = -\sum_{c_i\in{\mathcal{C}}}p_{c_i}\log p_{c_i},$ 
where the sum is taken over the set $\mathcal{C} = \{c_i\}$, for $i=1,\ldots,N$, consisting of symbols, which have a respective probability distribution $p_i = p(c_i)$ determining their occurrence. The amount of information contained by each symbol was determined by Shannon as  $\mathbb{I}(c_i) = -\log_2 p_i$, and satisfies the expression \cite{Gleiser:2018jpd}
\begin{equation}\label{ShannonInfo}
\langle \mathbb{I} \rangle = \sum_{c_i\in\mathcal C} p(c_i) \mathbb{I}(c_i) = -\sum_{c_i\in\mathcal C} p(c_i) \log_2 p(c_i) = S.
\end{equation}
The precise concept of information regards the lowest number of bits that are needed to 
convert symbols into a coded form, in such a way that the highest transmission rate of messages, that consist of an array of the given symbols, can be achieved across some transmission channel. Therefore, $\log_2 N$ is the number of bits to describe $N$ symbols.  
When one regards a uniform distribution, randomness is related to the probability distribution $p(c) = 1/N$, whose entropy is the highest possible at $\langle \mathbb{I} \rangle = \log_2 N$. 
The base of the logarithm to be used is a matter of choice, as the binary base 2 logarithm measures the information entropy
in bits, whereas one can employ the natural logarithm corresponding to measuring the information entropy in the natural unit of entropy (nat or Sh [shannon]). One nat is equal to 
$1/\ln 2$ shannon $\approx$ 1.4427 Sh. There is a correlation involving the CE and the Gibbs entropy, 
${S_{\scalebox{.66}{\textsc{thermo}}}=-k_{{\mathsf{B}} }\sum_a p_{a}\ln p_{a}\,}$, for $k_{\mathsf{B}}$ denoting the Boltzmann constant, and $p_i$ stands for the probability of occurrence of each specific detailed microscopic configuration of the system. Thermodynamic systems have microstates that do not have necessarily equal probabilities. 
The Boltzmann equation, 
$S_{\scalebox{.66}{\textsc{thermo}}}=k_{\mathsf{B}}\ln(\mathsf{W})$, 
where $\mathsf{W}$ is the number of equiprobable microstates, was generalized by the  Gibbs entropy, for the case of microstates that are not equiprobable. The CE quantifies the entropy in the context of information, as entropy in thermodynamics is equivalent to the portion of information, in the Shannon's sense, required for comprising microstates constituting a physical system \cite{Bernardini:2016hvx}. The spectral Fourier transform of the energy density, 
\begin{equation}
{\scalebox{.85}{\textsc{$\uprho$}}}(k)=\frac{1}{\sqrt{2\pi}}\int_{\mathbb{R}} e^{i{{k}}{\mathsf{r}}}\rho({\mathsf{r}})d\mathsf{r},
\label{collectivecoordinates}
\end{equation}
matches the collective coordinate concept in the statistical mechanics context, $\rho(r)\sim\sum e^{-i k_n r}{\scalebox{.85}{\textsc{$\uprho$}}}(k_n)$, when the limit of continuum mechanics is taken into account. 
The static structure factor is given by \beq
f(k_n)=\frac{\left\vert {\scalebox{.85}{\textsc{$\uprho$}}}(k_n)\right\vert ^{2}}{\sum_{a=1}^n\left\vert {\scalebox{.85}{\textsc{$\uprho$}}}(k_a)\right\vert^{2}},\eeq  which measures the probability distribution of any field pattern of the propagating waves with wavenumber $k_n$.

Going to the continuum mechanical limit, one can consider the DCE, which measures the information related to the energy density with $\ell^2$-norm. The Fourier transform of the energy density yields the energy spectral density, 
\begin{eqnarray}
{\scalebox{.85}{\textsc{$\uprho$}}}({k}) =\frac{1}{\sqrt{2\pi}} \int_{\mathbb{R}}  {\scalebox{.85}{\textsc{$\uprho$}}}({r})e^{-i{k}\cdot{r}}dr,\label{ftrans}
\end{eqnarray} and mimics collective coordinates in the continuum mechanical limit of statistical mechanics \cite{Bernardini:2016hvx}. The relative weight of distinct wave modes is quantified by the modal fraction,  
\begin{eqnarray}\label{modall}
\mathsf{f}_{\scalebox{.6}{\textsc{$\uprho$}}}(k) = \frac{|{\scalebox{.85}{\textsc{$\uprho$}}}({k})|^2}{\int_{\mathbb{R}}|{\scalebox{.85}{\textsc{$\uprho$}}}({\mathsf{k}})|^2d\mathsf{k}}.\label{modalf}
\end{eqnarray}
The modal fraction measures the weight carried by the wave modes associated with the wavenumber $k$. 
The portion of the information that is necessary to set out the shape of the density, here describing the scalar field coupled to gravity, is computed by the DCE
\cite{gs12b}, 
\begin{eqnarray}
S[{\scalebox{.85}{\textsc{$\uprho$}}}] = - \int_{\mathbb{R}} \invbreve{\mathsf{f}}_{{\scalebox{.6}{\textsc{$\uprho$}}}}({{\mathsf{k}}})\ln\left[\invbreve{\mathsf{f}}_{{\scalebox{.6}{\textsc{$\uprho$}}}}({\mathsf{k}})\right]d\mathsf{k},\label{ce1}
\end{eqnarray}
where $\invbreve{\mathsf{f}}_{\scalebox{.6}{\textsc{$\uprho$}}}({k})=\mathsf{f}_{\scalebox{.6}{\textsc{$\uprho$}}}({k})/\mathsf{f}_{\scalebox{.6}{\textsc{$\uprho$}}}^{\scalebox{.52}{\textsc{max}}}({k})$, being $\mathsf{f}_{\scalebox{.6}{\textsc{$\uprho$}}}^{\scalebox{.52}{\textsc{max}}}({k})$ the maximal value of the modal fraction at the wavenumber whose power spectral density has a peak.
The more the scalar field configuration is localized over the space, the more the configuration escalates across the momentum space. Consequently, the higher the DCE is.
Still, the configurational stability of the physical system here scrutinized and its regarded DCE are contiguous correlated quantities. The system stability regards the global minimum of the DCE, analyzed 
about the parameter that regulates the physical
features of the system. 

\subsection{Bondi mass and DCE for a 1-dimensional kink} 
The Bondi mass can be calculated as 
\be
M=4\pi\int_R^\infty \rho^{\Upphi}(r) r^2 dr,
\ee
leading us to consider for the Fourier transform (\ref{ftrans}) the linear energy density 
\be
\lambda(r)=4\pi r^2 \rho^\Upphi(r),\label{linear}
\ee
to derive the DCE (\ref{ce1}), $S[{\scalebox{.85}{\textsc{$\lambda$}}}]$, in the current setting.


\subsection{Numerical solver, pre-- and post--processing}
\clt{We use the well-established numerical solver known as the {\sc PITT} code to evolve a massless scalar field, adding new scripting for the DCE numerical calculations. Here the emphasis is not mainly on the gravitational critical  behavior, instead, it is in the DCE and how this quantity follows the main features of the gravitational collapse. Thus, the obtention of solutions is considered in this context as the input information, pre-processed. Therefore, the DCE calculation is the output information, post-processed. We resume now the numerical method to solve the Einstein--Klein--Gordon system, the critical behavior features for the kink, and some issues about the {\sc Python} scripting and using libraries.
\subsubsection{Pitt code}
Here we evolve the nonlinear massless scalar field using the 1D {\sc Pitt} code \cite{gwi92,gw92}. The algorithm adaptation to the present setting was developed in Ref. \cite{bglw96} and used in \cite{bcdrr16} to study the critical behavior as is summarized in the next subsection. 
The code is globally second-order accurate.
This code has been extended to get global energy conservation near the critical behavior in a setting as originally studied by Choptuik \cite{c93,Barreto:2017kta,b14}.}

\subsubsection{Critical behavior of a kink}
\clt{Ref. \cite{bcdrr16} studied the critical collapse of a kink leading to black hole formation and confirmed, for the
supercritical situation, a power-law scaling given by}
\be
M_{\scalebox{.66}{\textsc{BH}}} = M^{\scalebox{.66}{\textsc{crit}}}_{\scalebox{.66}{\textsc{BH}}} + K[\varepsilon-\varepsilon_{\scalebox{.66}{\textsc{crit}}}]^{2\gamma} + f\left(K[\varepsilon
-\varepsilon_{\scalebox{.66}{\textsc{crit}}}]^{2\gamma}\right),
\ee
with $\gamma\approx0.37$ and $M^{\scalebox{.66}{\textsc{crit}}}_{\scalebox{.66}{\textsc{BH}}}$, $K$, $\varepsilon_{\scalebox{.66}{\textsc{crit}}}$ depending on the initial data. For the subcritical case, the Bondi mass (in terms of a putative Bondi time) decays periodically in cascade with a $\Delta/2 \approx 1.7$ interval \cite{bcdrr16}, 
complying to the results in Ref. \cite{purrer}. It is interesting that in Ref. 
\cite{Santos-Olivan:2015yok}, in the context of the 
gravitational collapse of a massless scalar field for an 
asymptotically AdS spacetime, the critical exponent regulating  the 
multiple critical behaviors is close to twice $\gamma$, within 
$\approx 5\%$. It makes sense from the physical point of view since the mirror setting for the inner boundary in the kink 
case is akin to the reflecting boundary at infinity for the 
asymptotically AdS spacetime.  
\subsubsection{Scripting for DCE}
\clt{With the above challenge computations and extreme scenario as input, we developed an additional module using the standard {\sc Python} optimized libraries to get the DCE as output. A very simplified and efficient script uses a trapezoidal integration and the fast Fourier transform. The script was calibrated with different analytic models for signal processing and the output is convergent when the grid is progressively refined. For a typical run with $N_x=10^4$ grid points, each DCE value takes 2 seconds on a 1.8 GHz Dual-Core Intel Core i5, under OsX Big Sur. Doing sub-sampling in time and using extra scripting, we explore near the critical point bifurcations. Thus we can make a huge number of numerical experiments, minimizing the error of handling, processing the input data for the critical behavior, and especially for the black hole formation on the go of the evolution.}

\vspace{1cm}
\section{Two kinks}
\label{3s}

\subsection{Kink I: Globally perturbed static solution}
We use the static solution of Janis-Newman-Winicour as set in Ref. \cite{bglw96}. The static and asymptotically flat solution of the Einstein--Klein--Gordon system extremizes the energy functional, under the kink potential, which is similar to the  Minkowski solution
background. It can be derived in null coordinates, when the condition $\Upphi_{,u}=0$ is imposed in the equation of motion (\ref{we}), yielding
\beq
rV\Upphi_{,r}=\text{constant}.\label{cte}
\eeq
The solutions of Eq. (\ref{cte}) can be obtained when Eqs. (\ref{221}, \ref{222}) are taken into account to get rid of the dependence on the radial coordinate \cite{jnw68,bglw96}, and are expressed as
\begin{equation}
        \Upphi(V) = \frac{1}{4\sqrt{\pi}} \sec\!{\rm h}(\upalpha)\ln
        \left[\frac{R (e^{2\upalpha} - 1)+V}
         {R (e^{-2\upalpha} - 1)+V} \right],\label{eq:static}
\end{equation}
where {\blt{$V(r)$ has to be obtained numerically, inverting $r(V)\equiv r_\Upphi$, which reads} 
\begin{equation}
        r_\Upphi^2 =  e^{-4 \upalpha \tanh(\upalpha)}
               \frac{               [R (e^{2\upalpha} - 1)+V ]^{1 + \tanh(\upalpha)}}{[R (e^{-2\upalpha} - 1)+V]^{\tanh(\upalpha)-1}}.
                \label{eq:rstatic}
\end{equation}
Here the integration constant $\upalpha$ determines
the kink potential $A_\Upphi\equiv\Upphi(u,R)$, given by 
\begin{equation}
       A_\Upphi(\upalpha)=\frac{1}{4\sqrt{\pi}}\sec\!{\rm h}(\upalpha). \label{eq:kink}
\end{equation}
The spacetime has a naked singularity when analytically extended to
$r=0$~\cite{jnw68}. The Bondi mass associated with the  solution reads
\begin{equation}
    M_{\Upphi}(\upalpha)=2R\sinh^2(\upalpha) \exp\left[-2\upalpha\tanh(\upalpha)\right].
             \label{eq:statm}
\end{equation}
The static solutions were shown in Ref. \cite{bglw96} to consist of equilibrium configurations, with kink potential
$A_\Upphi(\upalpha)$ (\ref{eq:kink}) a mononotonically increasing function with respect to $\upalpha$, starting from  $A_\Upphi(0)=0$ up to a peak at the turning point $\upalpha_{\scalebox{.66}{\textsc{turn}}}\approxeq 1.199$, defined in such a \blt{way} that 
$\upalpha_{\scalebox{.66}{\textsc{turn}}}\tanh(\upalpha)_{\scalebox{.66}{\textsc{turn}}}=1$. In the range  $\upalpha>\upalpha_{\scalebox{.66}{\textsc{turn}}}$,   $A_\Upphi(\upalpha)$ is a 
monotonically decreasing function, with asymptotic behavior 
to $\lim_{\upalpha\rightarrow\infty}A_\Upphi(\upalpha)=0$. Hence, for kink potentials lower than  $A_\Upphi(\upalpha_{\scalebox{.66}{\textsc{turn}}})$,  two static equilibria exist, for each kink
amplitude. 
Thus the initial kink can be written as
\be
\Upphi(u=0,r;\upalpha)=\Upphi(\upalpha)+\frac{\varepsilon(1-2x)}{2r},\label{epsilon}
\ee
for $\varepsilon$ denoting a perturbation parameter, 
where $\Upphi(\upalpha)$ is the static solution and the compactified radial coordinate is defined as $x=r/(R+r)$. When $x=\frac12$ the boundary $r=R$ is attained, whereas $x=1$ corresponds to the future null infinity. 
 The stable equilibrium configurations can be therefore discussed when initial data evolve. Ref. \cite{bglw96} showed that the static equilibrium, $\Upphi(\upalpha)\vert_{\alpha\to1.42429}$ is unstable. At this phase, the system engages in dynamical phase transition, for low values of the perturbation parameter, to another stable state  $\Upphi(\upalpha)\vert_{\alpha\to1}$.

\subsection{Kink II: Far--from--equilibrium}
\blt{Now consider a kink potential, for which no static equilibria exist in the range $A>A_{\scalebox{.66}{\textsc{turn}}}$, with initial data
\beq\label{id}
\Upphi(u=0,r)=\frac{2(A_{\scalebox{.66}{\textsc{turn}}}+\varepsilon)}{r+R},
\eeq
where $A_{\scalebox{.66}{\textsc{turn}}}$ is the critical amplitude at the turning point. In this range, one expects initial states to go through a black hole end-stage of gravitational collapse. In particular regarding the initial data (\ref{id}), for the critical value of $\varepsilon=0$, there is no mass gap, and an infinite proper time would be needed to check the formation of a black hole.}

\section{Results and analysis}
\label{4s}

The DCE and the Bondi mass will be now  discussed with respect to the proper time and the Bondi time. The results of the computations are presented and discussed in what follows.

\begin{figure}
    \centering
    \includegraphics[width=.45\textwidth]{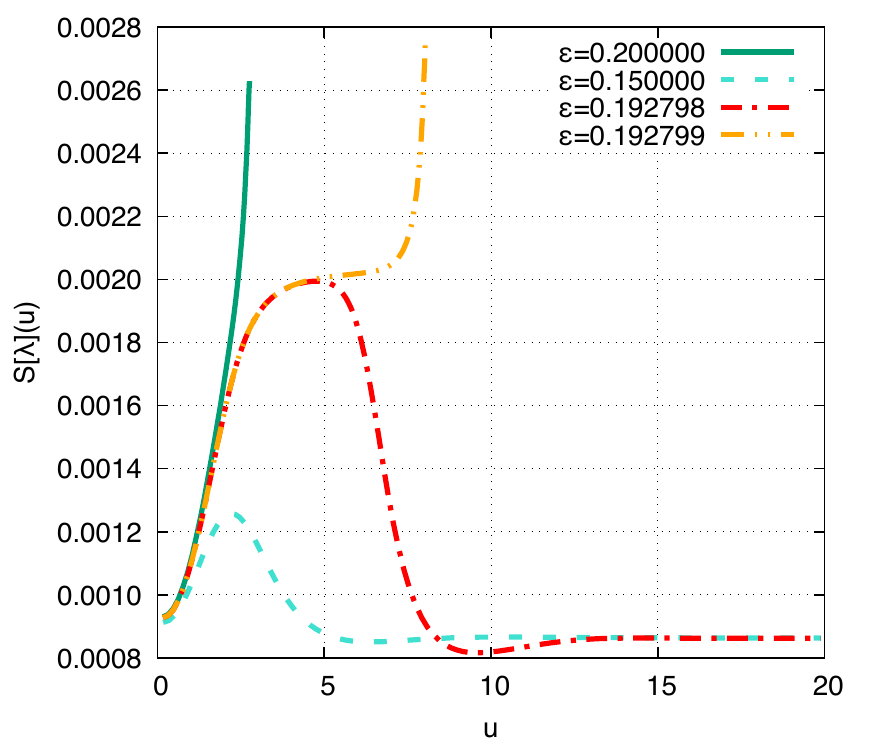}
    \includegraphics[width=.45\textwidth]{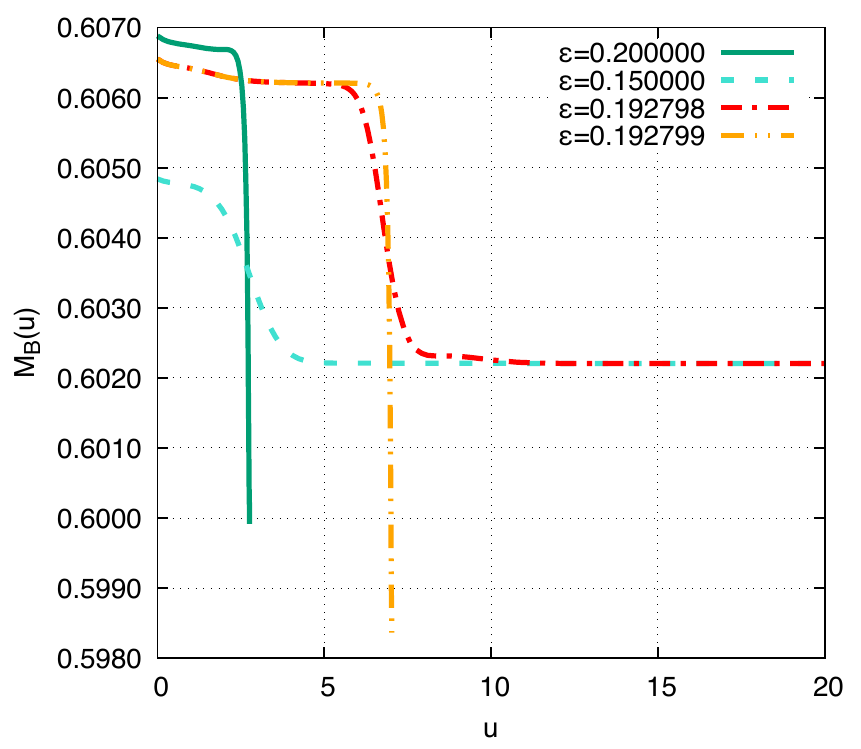}
    \caption{DCE (upper panel) and Bondi mass (lower panel) as a function of time, for the stable kink I ($\upalpha=1$). The existence of a critical amplitude perturbation $\varepsilon_{\scalebox{.66}{\textsc{crit$_{1}$}}}$ separates the evolutions between two final equilibria: a black hole or a static solution for $\upalpha=1$. It is a type-I phase transition, since it has a mass gap. It seems that very close to the critical solution the DCE holds a constant value. For a subcritical evolution the system evolves towards the same final DCE.}
    \label{fig:figure1}
\end{figure}

Fig. \ref{fig:figure1}, the plot on top, shows the DCE as a function of the proper time for the stable kink I. The black hole formation is evident whenever the perturbation parameter $\varepsilon$ in Eq. (\ref{epsilon}) surpasses the critical value $\varepsilon_{\scalebox{.66}{\textsc{crit$_{1}$}}}\approxeq0.1927985$, which determines a bifurcation point, wherefrom we can distinguish subcritical and supercritical evolution profiles. For subcritical evolution, regarding values of the perturbation parameter lower than $\varepsilon_{\scalebox{.66}{\textsc{crit$_{1}$}}}$, there is no black hole formation. In this way, the DCE exhibits a typical bifurcation for the critical gravitational collapse. In this case, with higher values of the proper time, the DCE  converges to the asymptotic preferential value of the static solution  recovering equilibria,  since it lacks a black hole collapse process. This critical value of the DCE, corresponding a  first-order phase transition, is invariant for all $\varepsilon<\varepsilon_{\scalebox{.66}{\textsc{crit$_{1}$}}}$. 
It is also worth emphasizing that for values $\varepsilon>\varepsilon_{\scalebox{.66}{\textsc{crit$_{1}$}}}$, corresponding to a supercritical evolution, the DCE is a monotonically increasing function of the proper time. 
The higher the value of the perturbation parameter $\varepsilon$ in Eq. (\ref{epsilon}), the higher the increment rate 
of the DCE with respect to the proper time is.
In the range $\varepsilon<\varepsilon_{\scalebox{.66}{\textsc{crit$_{1}$}}}$ that characterizes subcritical solutions, the DCE has a peak for any fixed value of the perturbation parameter. The higher the value of $\varepsilon$, the more the proper time elapses to attain the peak of the DCE. In the range $\varepsilon<\varepsilon_{\scalebox{.66}{\textsc{crit$_{1}$}}}$, the system returns to the state of equilibrium, $\Upphi(\upalpha)$, at late proper times. 
Besides, every fixed value of the perturbation parameter in the  range $\varepsilon<\varepsilon_{\scalebox{.66}{\textsc{crit$_{1}$}}}$ yields a global minimum of the DCE, corresponding to a transient plateau in the Bondi mass and to
          the maximum of the monopole, which decays up to a constant final value. The higher the value of $\varepsilon$ in this range, the more the proper time elapses to reach the respective minimum of the DCE and the lower the value of the minimum is. After the minimum, the DCE fastly converges to the asymptotic preferential value of the static solution. \blt{The location of global minima in different cases presents no scaling behavior.}

Now, Fig. \ref{fig:figure1}, the plot on the bottom, depicts the Bondi mass as a function of the proper time, for different values of the perturbation parameter $\varepsilon$ in Eq. (\ref{epsilon}). For the sake of consistency, the same values of the parameters in the plot on top are respectively employed. The first important feature regards the monotonicity of the Bondi mass with respect to the proper time. In the supercritical evolution range, the higher the value of the perturbation parameter $\varepsilon$, the steeper the Bondi mass decreases with respect to the proper time, up to the collapse process leading to black hole formation. The Bondi mass profile suddenly changes at $\varepsilon_{\scalebox{.66}{\textsc{crit$_{1}$}}}$ and in the subcritical evolution range the Bondi mass also converges to the asymptotic preferential value of the static solution for $\upalpha=1$,  with no black hole formation. When $\upalpha=1$, the kink potential assumes  $A_\Upphi(\upalpha)\vert_{\upalpha\to1}=0.36563$. The Bondi mass approaches its asymptotic equilibrium value, $M_\Upphi(\upalpha)\vert_{\upalpha\to1}\approxeq0.60220$. 
\begin{figure}
    \centering
    \includegraphics[width=0.46\textwidth]{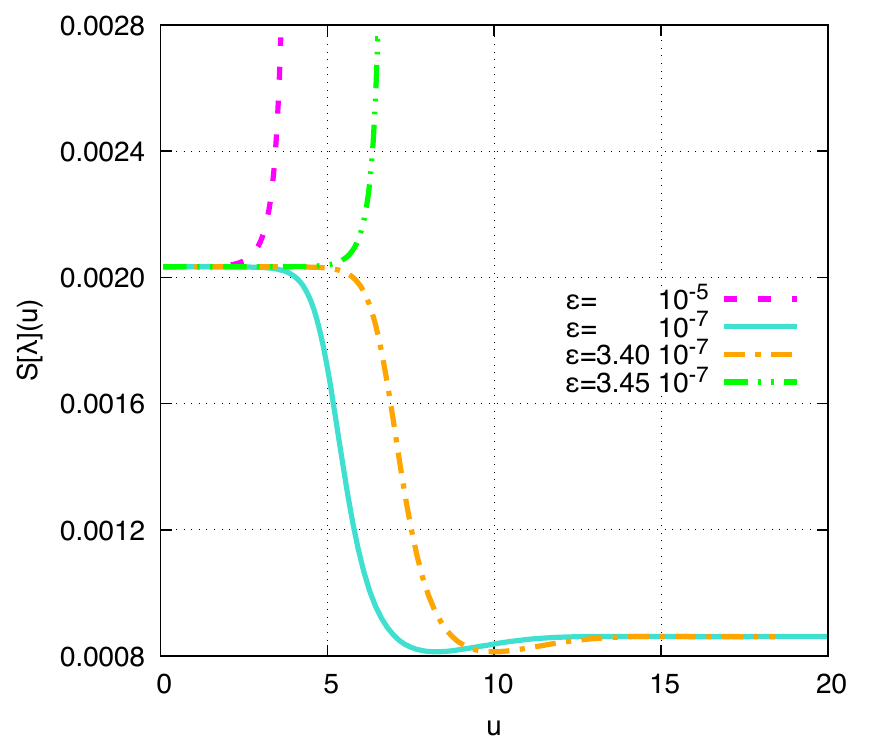}
    \includegraphics[width=0.46\textwidth]{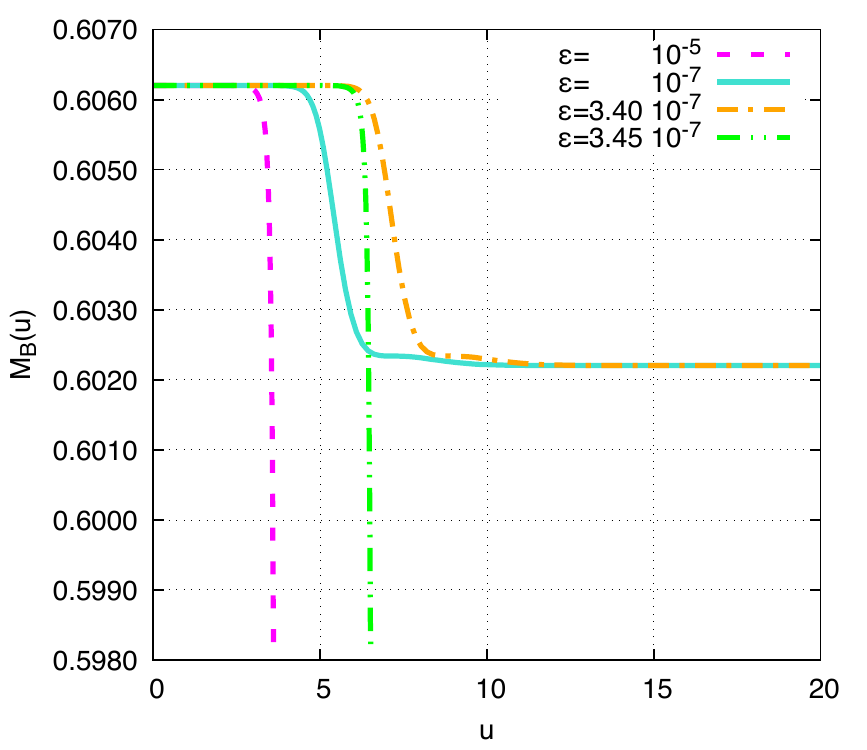}
    \caption{DCE (upper panel) and Bondi mass (lower panel) as a function of time, for the unstable kink I ($\upalpha=1.42429$). The existence of a critical amplitude perturbation $\varepsilon_{\scalebox{.66}{\textsc{crit$_{2}$}}}$ separates the  evolution  between two final equilibria: a black hole or a static solution for $\upalpha=1$. The phase transition is of first order, i.e. with a mass gap. It seems that very close to the critical solution the DCE holds a constant value. For a subcritical evolution, the system evolves towards the same final DCE. For a supercritical evolution, the system seems to evolve towards the same final DCE.}
    \label{fig:figure2}
\end{figure}

In Fig. \ref{fig:figure2} the unstable kink I is illustrated and analyzed. The upper plot depicts  the DCE as a function of the proper time for the unstable kink I ($\upalpha=1.42429$). The black hole formation is again manifest whenever the perturbation parameter $\varepsilon$ in Eq. (\ref{epsilon}) reaches values that are higher than the critical value $\varepsilon_{\scalebox{.66}{\textsc{crit$_{2}$}}}\approxeq 3.42\times 10^{-7}$,  which regulates a bifurcation point for the unstable kink I. Values $\varepsilon\gtrless\varepsilon_{\scalebox{.66}{\textsc{crit$_{2}$}}}$ respectively determine supercritical and subcritical evolution of the gravitational kink. Similarly to the stable kink I, the range  $\varepsilon<\varepsilon_{\scalebox{.66}{\textsc{crit$_{2}$}}}$ yields no black hole formation. Also, the DCE  converges to the asymptotic preferential value of the static solution that regards  a  first-order phase transition, being invariant for all  $\varepsilon<\varepsilon_{\scalebox{.66}{\textsc{crit$_{2}$}}}$ in the subcritical evolution profile. Every fixed value of the perturbation parameter in the range $\varepsilon<\varepsilon_{\scalebox{.66}{\textsc{crit$_{2}$}}}$ yields a global minimum of the DCE. The lower the value of $\varepsilon$ in this range, the less the proper time elapses to reach the respective minimum of the DCE and the higher the value of the minimum is. After the minimum, the DCE rapidly converges to the asymptotic preferential value of the static solution for $\upalpha=1$.
However, in the supercritical evolution range  $\varepsilon>\varepsilon_{\scalebox{.66}{\textsc{crit$_{2}$}}}$,  the DCE is a monotonically non-decreasing function of the proper time. 
The lower the value of  $\varepsilon$ in Eq. (\ref{epsilon}), the lower the increment rate 
of the DCE, with respect to the proper time, is.

The plot on the bottom in Fig. \ref{fig:figure2}  depicts the Bondi mass as a function of the proper time, for different values of the perturbation parameter $\varepsilon$. For the sake of consistency, the same values of the parameters in the plot on top in Fig. \ref{fig:figure2} are respectively employed. The Bondi mass is a monotonically non-increasing function of the proper time. In the supercritical evolution range, the higher the value of the perturbation parameter $\varepsilon$, the sharper the Bondi mass decreases, up to the collapse process leading to black hole formation. The Bondi mass profile is then modified at the bifurcation point defined by the critical value $\varepsilon_{\scalebox{.66}{\textsc{crit$_{2}$}}}$. In the subcritical evolution range, the Bondi mass also converges to the asymptotic preferential value of the static solution for $\upalpha=1$, when black holes are not formed. 

\begin{figure}
    \centering
    \includegraphics[width=0.45\textwidth]{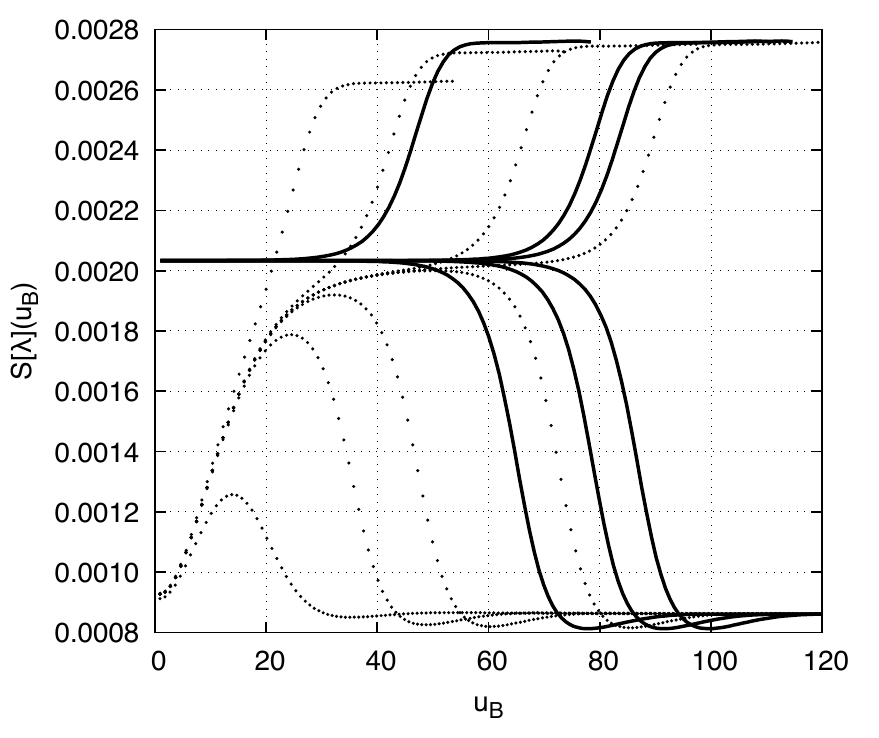}
    \caption{DCE as a function of the Bondi time,  for several kink I evolution profiles: stable (dotted lines) and unstable (continuous lines). }
    \label{fig:figure3}
\end{figure}

Fig.  \ref{fig:figure3} illustrates the DCE as a function of the Bondi time,  for several kink I evolution profiles, including stable and unstable configurations. 
The critical, subcritical, and supercritical asymptotic values of the DCE are the same for evolution profiles in the respective channel. 
As expected, there is a subtle qualitative difference when comparing to 
the DCE, here as a function of the Bondi time, and in 
Fig. \ref{fig:figure1}, as a function of the proper time, 
mainly in what concerns the supercritical evolution profile.
Although being still a monotonically non-decreasing function of the Bondi time, the DCE at the supercritical evolution profile
has a plateau at the preferential asymptotic value. There are also bifurcation points, from which subcritical and supercritical evolution profiles can be discerned. Yet the  subcritical evolution points to  no black hole origination, and  the DCE evinces a typical bifurcation for the critical gravitational collapse. For higher values of the Bondi time, the DCE  converges to the asymptotic preferential value of the static solution, for both the stable and unstable evolution profiles. 
The higher the value of the perturbation parameter, the higher the increment rate of the DCE with respect to the Bondi time is.
Subcritical solutions present a peak of the DCE. The higher the value of the perturbation parameter, the more the Bondi time elapses to reach the peak of the DCE.
Also, in the subcritical range, there are global minima of the DCE for each fixed value of the perturbation parameter. The lower the value of the perturbation parameter in the subcritical range, the less the proper time elapses to reach the respective global minimum of the DCE, and the higher the value of the global minimum is. For values of the Bondi time that are in the future when compared to the Bondi time regarding the global minima, the DCE fastly converges to the asymptotic preferential value of the static solution. \blt{In the subcritical range $\varepsilon<\varepsilon_{\scalebox{.66}{\textsc{crit$_{1}$}}}$, we numerically verified that for the DCE as a function of the Bondi time, the maximum value of DCE, denoted by $S_{\scalebox{.52}{\textsc{peak}}}$,  scales with $\varepsilon$ (see Fig. \ref{fig:figure3}) as 
\beq
\!\!\!\!\!\!\!\!\!S_{\scalebox{.52}{\textsc{peak}}}(\varepsilon)&=&13013.5038 \varepsilon^3 -6950.1781 \varepsilon^2+1229.6481 \varepsilon\nonumber\\&&\qquad\qquad\qquad\qquad\quad\qquad\qquad-71.9875,
\eeq
within $0.01\%$ numerical error.}
\begin{figure}
    \centering
    \includegraphics[width=0.45\textwidth]{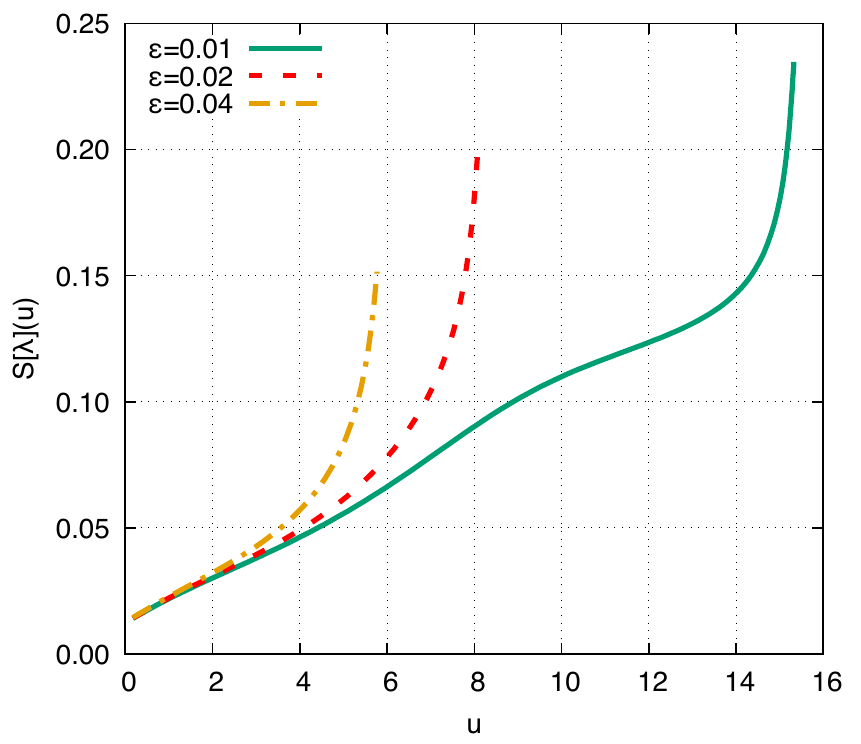}
    \includegraphics[width=0.45\textwidth]{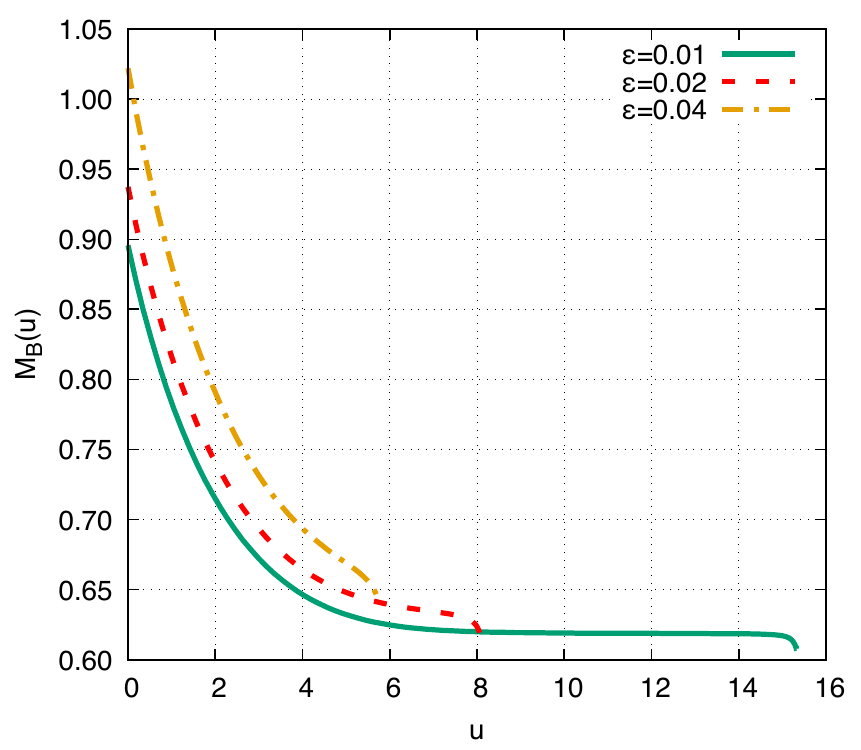}
    \caption{DCE (multiplied by $10^2$; upper panel) and Bondi mass (lower panel) as a function of time, for the kink II. Always a black hole is formed.}
    \label{fig:figure4}
\end{figure}
The plot on top in Fig.  \ref{fig:figure4} refers to the DCE as a function of the proper time for the kink II. Irrespectively of the value of the perturbation parameter, the DCE is a monotonically increasing function of the proper time. For values $\varepsilon\gtrsim 0.015$, there is an interesting change of behavior of the DCE characterized by ${d^2S[\lambda](u)}/{du^2}>0$, for all values of the proper time. Still, for lower values of $\varepsilon$, ${d^2S[\lambda](u)}/{du^2}>0$ just for higher values of the proper time. It is worth emphasizing that the higher the value of the perturbation parameter $\varepsilon$, the steeper the increment rate of the DCE is, respectively with higher values of  ${d^2S[\lambda](u)}/{du^2}$, accordingly. In all cases analyzed, the final state of the kink is the gravitational collapse, leading to black hole formation as the scalar field pours its associated hair. The Bondi mass is also presented as a function of the proper time for the kink II, in the plot on the bottom in Fig.  \ref{fig:figure4}. The Bondi mass is a monotonically decreasing function of the proper time, presenting ${d^2S[\lambda](u)}/{du^2}>0$  in almost all the ranges respectively illustrated, but an end-stage corresponding to a very short proper time interval, where the Bondi mass has a sudden decay to a final state. This attribute holds for any value of the perturbation parameter. The higher the value of the perturbation parameter, the shorter the proper time elapses for the Bondi mass to reach the end-stage.




\section{Concluding remarks}
\label{conclu}

\clt{The gravitational collapse of two types of kinks was heretofore 
investigated, employing the DCE setup. Globally perturbed static kinks, prospected here in both the stable and unstable cases, source two end-stage equilibria, consisting of either a static solution or the formation of the black hole near the scalar field critical collapse. This ramification, determined by a  critical amplitude perturbation, categorizes a
first-order phase transition containing a mass gap, wherein  the black hole formation represents a supercritical domain, whereas static solutions regard the subcritical range, with no possibility of forming a black hole by the scalar field collapse. The DCE precisely designates the critical amplitude perturbation as a branching process governing the critical gravitational collapse, also emulating the respective supercritical and subcritical regions for the associated Bondi mass.   In the subcritical case, both the Bondi mass and the DCE have been shown to converge, respectively, to the asymptotic preferential values of static solutions. For the far-from-equilibrium kink solutions, the DCE corroborates black hole formation from the gravitational collapse of the kink, for all cases of the critical amplitude perturbation. Regardless of the value of the perturbation parameter, the DCE was derived as a monotonically increasing function of the proper time, diverging at finite proper time elapse, leading to collapse into a black hole. On the other hand, the Bondi mass associated with far-from-equilibrium kinks always decreases with time. Black holes formed by the process here described and endorsed by DCE methods might exist as an observable physical by-product of the evolution of solutions of the Einstein--Klein--Gordon system. They may form coalescing black hole binary mergers emitting gravitational waves radiation, whose physical signatures can be predominantly probed with the gravitational wave detectors \cite{Baker:2005vv}. 
Their observational analysis 
may provide a groundbreaking source of information about the strong-field regime of general relativity. The analysis of Extreme Mass Ratio Inspirals can impart useful hints about Einstein--Klein--Gordon systems, since perturbative models of the inspiral sector record a noteworthy imprint of a scalar field on observable gravitational wave radiation \cite{Maselli:2020zgv}. \blt{
It is worth mentioning that the
leading-order influence arises from dipolar radiation emission, with a scalar field coupled to gravity. At the leading order, each black hole merger in the binary system emulates a monopole, whereas the binary system emits dipole radiation, also imposing bounds on massless scalar fields according to the latest binary pulsar observations \cite{Freire:2012mg}. Extreme mass-ratio inspirals onto supermassive black holes play a prominent role in the phenomenological aspects here studied, as scalar monopoles are accelerated  by gravity at the supermassive black hole, emitting scalar radiation that comprises mostly of dipole radiation. The growing  effect at the long-lasting 
inspiral can be detectable at ground-based detectors \cite{LIGOScientific:2017vwq,Xue:2019nlf}, with additional energy loss, and subsequent dephasing of the gravitational waveform, depending on the scalar field as here proposed, dressing the supermassive black hole. }}

\medbreak
\subsubsection*{Acknowledgments} 
RdR~is grateful to FAPESP (Grants No. 2017/18897-8, No. 2021/01089-1, and No. 2022/01734-7) and the National Council for Scientific and Technological Development -- CNPq (Grants No. 303390/2019-0 and No. 402535/2021-9), for partial financial support.


\end{document}